\RequirePackage{ifpdf}
\ifpdf 
\documentclass[pdftex]{sigma}
\else
\documentclass{sigma}
\fi

\begin{document}
\allowdisplaybreaks

\renewcommand{\PaperNumber}{050}

\FirstPageHeading

\ShortArticleName{Twin ``Fano-Snowf\/lakes'' over the  Smallest Ring of
Ternions}

\ArticleName{Twin ``Fano-Snowf\/lakes''\\ over the  Smallest Ring of
Ternions}

\Author{Metod SANIGA~$^\dag$, Hans HAVLICEK~$^\ddag$, Michel PLANAT~$^\S$ and Petr PRACNA~$^\P$}

\AuthorNameForHeading{M. Saniga, H. Havlicek, M. Planat and P. Pracna}

\Address{$^\dag$~Astronomical Institute, Slovak Academy of Sciences,\\
$\phantom{^\dag}$~SK-05960 Tatransk\' a Lomnica, Slovak Republic}
\EmailD{\href{mailto:msaniga@astro.sk}{msaniga@astro.sk}}
\URLaddressD{\url{http://www.ta3.sk/~msaniga/}}

\Address{$^\ddag$~Institut f\"ur Diskrete Mathematik und Geometrie,
Technische Universit\" at Wien,\\
$\phantom{^\ddag}$~Wiedner Hauptstra\ss e 8--10, A-1040 Vienna, Austria}
\EmailD{\href{mailto:havlicek@geometrie.tuwien.ac.at}{havlicek@geometrie.tuwien.ac.at}}

\Address{$^\S$~Institut FEMTO-ST/CNRS, MN2S,
32 Avenue de
l'Observatoire,\\
$\phantom{^\S}$~F-25044 Besan\c con Cedex, France}
\EmailD{\href{mailto:michel.planat@femto-st.fr}{michel.planat@femto-st.fr}}

\Address{$^\P$~J.\ Heyrovsk\' y Institute of Physical
Chemistry, v.v.i., Academy of Sciences of the Czech Republic,\\
$\phantom{^\P}$~Dolej\v skova 3, CZ-182 23 Prague 8, Czech Republic}
\EmailD{\href{mailto:pracna@jh-inst.cas.cz}{pracna@jh-inst.cas.cz}}

\ArticleDates{Received May 02, 2008, in f\/inal form May 30,
2008; Published online June 04, 2008}

\Abstract{Given a f\/inite associative ring with unity, $R$, any
free (left) cyclic submodule (FCS) generated by a \textit{uni}modular
$(n+1)$-tuple of elements of $R$ represents a point of the
$n$-dimensional projective space over $R$. Suppose that $R$ also
features FCSs generated by $(n+1)$-tuples that are \textit{not}
unimodular: what kind of geometry can be ascribed to such FCSs?
Here, we (partially) answer this question for $n=2$ when $R$ is
the (unique) non-commutative ring of order eight. The
corresponding geometry is dubbed a ``Fano-Snowf\/lake'' due to its
diagrammatic appearance and the fact that it contains the Fano
plane in its center. There exist, in fact, two such conf\/igurations
-- each being tied to either of the two maximal ideals of the
ring -- which have the Fano plane in common and can, therefore,
be viewed as twins. Potential relevance of these
noteworthy conf\/igurations to quantum information theory and
stringy black holes is also outlined.}


\Classification{51C05; 51Exx}

\looseness=1
Classical projective geometries, i.e.,
the ones where coordinates are elements of a (possibly skew)
f\/ield, represent a venerable and well-developed branch of
algebraic geometry. On the contrary, geometries over rings which are not f\/ields,
despite a quite abundant literature \cite{tv}, still lie well outside
the current mainstream research and there is still a lot to be
done and disco\-vered. Our interest in these geometries stems from
the recent recognition of their importance for the f\/ield of
quantum information theory (see, e.g., \cite{psk,spph,hs07,pb,hs08,spp} and references
therein). Being motivated by this fact, we have had a detailed
look at the f\/ine structure of projective lines and, to a lesser
extent, projective planes over a number of f\/inite (both
commutative and non-commutative) rings up to order 31. In doing so, we came
across a number of interesting features \cite{san} which have apparently
not been treated so far by either mathematicians or physicists.
This short note aims at acquainting the reader with, in our
opinion, most fascinating of them.

\newpage

     Let us consider an associative ring with unity $1$ $(\neq 0)$, $R$, and denote the
left module on $n+1$ generators over $R$  by $R^{n+1}$. The set
$R(r_1, r_2,\dots, r_{n+1})$, def\/ined as follows
\[
R(r_1, r_2,\dots, r_{n+1}):= \left\{ (\alpha r_1, \alpha r_2,\dots, \alpha  r_{n+1})\, |\, (r_1, r_2,\dots, r_{n+1}) \in R^{n+1}, \ \alpha \in R \right\},
\]
is a {\it left} cyclic submodule of $ R^{n+1}$. Any such submodule
is called {\it free} if the mapping $\alpha \mapsto (\alpha r_1,
\alpha r_2,\dots, \alpha  r_{n+1}) $ is injective, i.e., if
$(\alpha r_1, \alpha r_2,\dots, \alpha  r_{n+1}) $ are all {\it
distinct}. Next, we shall call $(r_1, r_2,\dots, r_{n+1}) \in
R^{n+1}$ {\it uni}modular if there exist elements $x_1,
x_2,\dots, x_{n+1}$ in $R$ such that
\[
r_1 x_1 + r_2 x_2 + \cdots + r_{n+1} x_{n+1} = 1.
\]
It is a very well-known fact (see, e.g., \cite{v81,v95,her}) that if $(r_1,
r_2,\dots, r_{n+1})$ is unimodular, then $R(r_1, r_2,\dots,
r_{n+1})$ is free; any such free cyclic submodule represents a
point of the $n$-dimensional projective space def\/ined over $R$~\cite{v95}. The converse statement, however, is not generally true. That
is, there exist rings which also give rise to free cyclic
submodules featuring exclusively non-unimodular ($n+1$)-tuples,
i.e., free cyclic submodules that cannot be associated with any
point of the corresponding projective space.  The f\/irst case when
this happens is the unique non-commutative ring of order eight,
$R_{\diamondsuit}$, i.e., the ring isomorphic to the one of
upper (or lower) triangular two-by-two matrices over the Galois f\/ield of two
elements.  Let us therefore have a~detailed look at this case.

To this end in view, we f\/irst introduce the standard matrix representation
of $R_{\diamondsuit}$ \cite{benz,lpw},
\[
   R_{\diamondsuit}  \equiv \left\{ \left(
\begin{array}{cc}
a & b \\
0 & c \\
\end{array}
\right) \mid ~ a, b, c \in GF(2) \right\},
\]
from where it is readily seen that the ring contains two maximal (two-sided) ideals,
\[
   I_1 =   \left\{ \left(
\begin{array}{cc}
0 & b \\
0 & c \\
\end{array}
\right) \mid ~  b, c \in GF(2) \right\}
\]
and
\[
   I_2 = \left\{ \left(
\begin{array}{cc}
a & b \\
0 & 0 \\
\end{array}
\right) \mid ~ a, b \in GF(2) \right\},
\]
which give rise to a non-trivial (two-sided) Jacobson radical $J$,
\[
   J = I_1 \cap I_2 = \left\{ \left(
\begin{array}{cc}
0 & b \\
0 & 0 \\
\end{array}
\right) \mid ~ b \in GF(2) \right\}.
\]
As for our further purposes it will be more convenient to work
with numbers than matrices, we shall relabel the elements of
$R_{\diamondsuit}$ as follows
\begin{gather}
0 \equiv \left(
\begin{array}{cc}
0 & 0 \\
0 & 0
\end{array}
\right),\qquad 1 \equiv \left(
\begin{array}{cc}
1 & 0 \\
0 & 1
\end{array}
\right),\qquad 2 \equiv \left(
\begin{array}{cc}
1 & 1 \\
0 & 1
\end{array}
\right),\qquad
3 \equiv \left(
\begin{array}{cc}
1 & 1 \\
0 & 0 \\
\end{array}
\right), \nonumber \\
4 \equiv \left(
\begin{array}{cc}
0 & 0 \\
0 & 1
\end{array}
\right),\qquad 5 \equiv \left(
\begin{array}{cc}
1 & 0 \\
0 & 0 \\
\end{array}
\right),\qquad  6 \equiv \left(
\begin{array}{cc}
0 & 1 \\
0 & 0 \\
\end{array}
\right),\qquad
7 \equiv \left(
\begin{array}{cc}
0 & 1 \\
0 & 1 \\
\end{array}
\right).\label{eq7}
\end{gather}
The addition and multiplication in the ring is that of matrices over $GF(2)$, which in our compact notation
reads as shown in Table~\ref{table1} (adopted, with a slight notational dif\/ference, from \cite{noe}).
From the multiplication table (as well as from expressions~\eqref{eq7}) it follows that,
apart from unity, there is only one more invertible element, 2, and that the zero-divisors are of two kinds:
{\it nil}potent (0~and~6) and {\it idem}potent (3, 4, 5, and 7). The two maximal ideals now acquire the form
\[
I_1 : = \left\{0, 4, 6, 7 \right\}
\qquad \mbox{and} \qquad
I_2 : = \left\{0, 3, 5, 6 \right\},
\]
and the Jacobson radical reads,
\[
J = I_1 \cap I_2 = \left\{0, 6 \right\}.
\]
Conf\/ining to the case of $n = 2$, we will now demonstrate that there exist triples of elements $(r_1, r_2, r_3) \in R_{\diamondsuit}^{3}$ which
are not unimodular and yet they generate free cyclic submodules.

\begin{table}[t]
\centering
\caption{Addition ({\it left}) and multiplication ({\it right}) in
$R_{\diamondsuit}$.}\label{table1} \vspace*{0.2cm}
\begin{tabular}{||c|cccccccc||}
\hline \hline
$+$ & 0 & 1 & 2 & 3 & 4 & 5 & 6 & 7 \\
\hline
0 & 0 & 1 & 2 & 3 & 4 & 5 & 6 & 7 \\
1 & 1 & 0 & 6 & 7 & 5 & 4 & 2 & 3 \\
2 & 2 & 6 & 0 & 4 & 3 & 7 & 1 & 5 \\
3 & 3 & 7 & 4 & 0 & 2 & 6 & 5 & 1 \\
4 & 4 & 5 & 3 & 2 & 0 & 1 & 7 & 6 \\
5 & 5 & 4 & 7 & 6 & 1 & 0 & 3 & 2 \\
6 & 6 & 2 & 1 & 5 & 7 & 3 & 0 & 4 \\
7 & 7 & 3 & 5 & 1 & 6 & 2 & 4 & 0 \\
\hline \hline
\end{tabular}~~~~~
\begin{tabular}{||c|cccccccc||}
\hline \hline
$\times$ & 0 & 1 & 2 & 3 & 4 & 5 & 6 & 7  \\
\hline
0 &  0 & 0 & 0 & 0 & 0 & 0 & 0 & 0 \\
1 &  0 & 1 & 2 & 3 & 4 & 5 & 6 & 7 \\
2 &  0 & 2 & 1 & 3 & 7 & 5 & 6 & 4 \\
3 &  0 & 3 & 5 & 3 & 6 & 5 & 6 & 0 \\
4 &  0 & 4 & 4 & 0 & 4 & 0 & 0 & 4 \\
5 &  0 & 5 & 3 & 3 & 0 & 5 & 6 & 6 \\
6 &  0 & 6 & 6 & 0 & 6 & 0 & 0 & 6 \\
7 &  0 & 7 & 7 & 0 & 7 & 0 & 0 & 7 \\
\hline \hline
\end{tabular}
\end{table}

\begin{figure}[t]
\centerline{\includegraphics[width=7.5truecm]{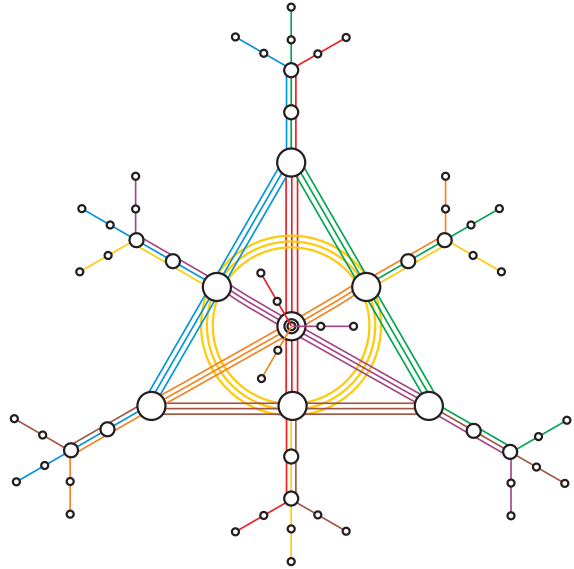}}
\caption{The ``Fano-Snowf\/lake'' -- a diagrammatic
illustration of a very intricate relation between the 21 free left
cyclic submodules generated by non-unimodular triples of
$R_{\diamondsuit}^{3}$ (represented here by the circles of
dif\/ferent sizes as explained in the text). As the $(0,0,0)$ triple
is not shown, each submodule is represented by seven circles
(three big, two medium-sized and two small) lying on a common
segmented/broken line. The colors were chosen in such a way to
make (the lines of) the Fano plane sitting in the middle of the
conf\/iguration readily discernible. There are seven ``protrusions'',
or ``extensions'', of the Fano plane emanating from its points,
each giving rise to three more ramif\/ications. (The
``protrusion/extension'' emanating from the point at the very
center is illustrated as going perpendicularly out of the sheet of
page so that only its three ramif\/ications can be seen properly.)
On a side note, any of the seven sets of three submodules represented
by the same color stands for an $n = 1$ case.}\label{fig1}
\end{figure}

It is a straightforward task to verify that such triples are tied uniquely to
the {\it first} maximal ideal ($I_1$); here, we shall list them implicitly together with the corresponding free cyclic submodules:
\begin{gather*}
 R_{\diamondsuit}(4,6,7)=R_{\diamondsuit}(7,6,4) \\
 \phantom{R_{\diamondsuit}(4,6,7)}{} = \left\{(0,0,0), (4,6,7), (7,6,4), (6,6,0), (4,0,4), (0,6,6), (6,0,6), (7,0,7) \right\} ,
\\
R_{\diamondsuit}(4,7,6)=R_{\diamondsuit}(7,4,6)\\
\phantom{R_{\diamondsuit}(4,7,6)}{} = \left\{(0,0,0), (4,7,6), (7,4,6), (6,0,6), (4,4,0), (0,6,6), (6,6,0), (7,7,0) \right\},\\
R_{\diamondsuit}(6,4,7)=R_{\diamondsuit}(6,7,4)\\
 \phantom{R_{\diamondsuit}(6,4,7)}{} = \left\{(0,0,0), (6,4,7), (6,7,4), (6,6,0), (0,4,4), (6,0,6), (0,6,6), (0,7,7) \right\},\\
R_{\diamondsuit}(4,4,7)=R_{\diamondsuit}(7,7,4)\\
 \phantom{R_{\diamondsuit}(4,4,7)}{} = \left\{(0,0,0), (4,4,7), (7,7,4), (6,6,0), (4,4,4), (0,0,6), (6,6,6), (7,7,7) \right\},\\
R_{\diamondsuit}(4,7,4)=R_{\diamondsuit}(7,4,7) \\
\phantom{R_{\diamondsuit}(4,7,4)}{} = \left\{(0,0,0), (4,7,4), (7,4,7), (6,0,6), (4,4,4), (0,6,0), (6,6,6), (7,7,7) \right\},\\
R_{\diamondsuit}(7,4,4)=R_{\diamondsuit}(4,7,7) \\
\phantom{R_{\diamondsuit}(7,4,4)}{} = \left\{(0,0,0), (7,4,4), (4,7,7), (0,6,6), (4,4,4), (6,0,0), (6,6,6), (7,7,7) \right\},\\
R_{\diamondsuit}(4,4,6)=R_{\diamondsuit}(7,7,6) \\
\phantom{R_{\diamondsuit}(4,4,6)}{} = \left\{(0,0,0), (4,4,6), (7,7,6), (6,6,6), (4,4,0), (0,0,6), (6,6,0), (7,7,0) \right\},\\
R_{\diamondsuit}(4,6,4)=R_{\diamondsuit}(7,6,7)\\
 \phantom{R_{\diamondsuit}(4,6,4)}{} = \left\{(0,0,0), (4,6,4), (7,6,7), (6,6,6), (4,0,4), (0,6,0), (6,0,6), (7,0,7) \right\},\\
R_{\diamondsuit}(6,4,4)=R_{\diamondsuit}(6,7,7) \\
\phantom{R_{\diamondsuit}(6,4,4)}{} = \left\{(0,0,0), (6,4,4), (6,7,7), (6,6,6), (0,4,4), (6,0,0), (0,6,6), (0,7,7) \right\},\\
R_{\diamondsuit}(6,6,7)=R_{\diamondsuit}(6,6,4)\\
\phantom{R_{\diamondsuit}(6,6,7)}{} = \left\{(0,0,0), (6,6,7), (6,6,4), (6,6,0), (0,0,4), (6,6,6), (0,0,6), (0,0,7) \right\},\\
R_{\diamondsuit}(6,7,6)=R_{\diamondsuit}(6,4,6) \\
\phantom{R_{\diamondsuit}(6,7,6)}{} = \left\{(0,0,0), (6,7,6), (6,4,6), (6,0,6), (0,4,0), (6,6,6), (0,6,0), (0,7,0) \right\},\\
R_{\diamondsuit}(7,6,6)=R_{\diamondsuit}(4,6,6)\\
\phantom{R_{\diamondsuit}(6,7,6)}{} = \left\{(0,0,0), (7,6,6), (4,6,6), (0,6,6), (4,0,0), (6,6,6), (6,0,0), (7,0,0) \right\},\\
R_{\diamondsuit}(0,6,7)=R_{\diamondsuit}(0,6,4) \\
\phantom{R_{\diamondsuit}(6,7,6)}{}= \left\{(0,0,0), (0,6,7), (0,6,4), (0,6,0), (0,0,4), (0,6,6), (0,0,6), (0,0,7) \right\},\\
R_{\diamondsuit}(0,7,6)=R_{\diamondsuit}(0,4,6) \\
\phantom{R_{\diamondsuit}(6,7,6)}{}= \left\{(0,0,0), (0,7,6), (0,4,6), (0,0,6), (0,4,0), (0,6,6), (0,6,0), (0,7,0) \right\},\\
R_{\diamondsuit}(0,4,7)=R_{\diamondsuit}(0,7,4) \\
\phantom{R_{\diamondsuit}(6,7,6)}{}= \left\{(0,0,0), (0,4,7), (0,7,4), (0,6,0), (0,4,4), (0,0,6), (0,6,6), (0,7,7) \right\},\\
R_{\diamondsuit}(6,0,7)=R_{\diamondsuit}(6,0,4) \\
\phantom{R_{\diamondsuit}(6,7,6)}{}= \left\{(0,0,0), (6,0,7), (6,0,4), (6,0,0), (0,0,4), (6,0,6), (0,0,6), (0,0,7) \right\},\\
R_{\diamondsuit}(7,0,6)=R_{\diamondsuit}(4,0,6)\\
\phantom{R_{\diamondsuit}(6,7,6)}{} = \left\{(0,0,0), (7,0,6), (4,0,6), (0,0,6), (4,0,0), (6,0,6), (6,0,0), (7,0,0) \right\},\\
R_{\diamondsuit}(4,0,7)=R_{\diamondsuit}(7,0,4)\\
\phantom{R_{\diamondsuit}(6,7,6)}{} = \left\{(0,0,0), (4,0,7), (7,0,4), (6,0,0), (4,0,4), (0,0,6), (6,0,6), (7,0,7) \right\},\\
R_{\diamondsuit}(6,7,0)=R_{\diamondsuit}(6,4,0) \\
\phantom{R_{\diamondsuit}(6,7,6)}{}= \left\{(0,0,0), (6,7,0), (6,4,0), (6,0,0), (0,4,0), (6,6,0), (0,6,0), (0,7,0) \right\},\\
R_{\diamondsuit}(7,6,0)=R_{\diamondsuit}(4,6,0) \\
\phantom{R_{\diamondsuit}(6,7,6)}{}= \left\{(0,0,0), (7,6,0), (4,6,0), (0,6,0), (4,0,0), (6,6,0), (6,0,0), (7,0,0) \right\},\\
R_{\diamondsuit}(4,7,0)=R_{\diamondsuit}(7,4,0)\\
\phantom{R_{\diamondsuit}(6,7,6)}{} = \left\{(0,0,0), (4,7,0), (7,4,0), (6,0,0), (4,4,0), (0,6,0), (6,6,0), (7,7,0) \right\}.
 \end{gather*}
We f\/ind altogether 42 non-unimodular triples of elements of $
R_{\diamondsuit}^{3}$ generating 21 distinct free left cyclic
submodules which in their entirety comprise all 64 triples of elements formed from $I_1$. These submodules are, as illustrated in Fig.~\ref{fig1}, very
intricately ``interwoven'' with each other as:
\begin{itemize}\itemsep=0pt

\item   the triple $(0,0,0)$ (not shown in the f\/igure) is the common meet of all of them;

\item   each of the seven triples $(6,0,0)$, $(0,6,0)$, $(0,0,6)$, $(6,6,0)$, $(6,0,6)$, $(0,6,6)$, $(6,6,6)$ (represented
by big circles in the f\/igure) shares nine of them;

\item   each of the 14 triples $(4,0,0)$, $(0,4,0)$, $(0,0,4)$, $(4,4,0)$, $(4,0,4)$, $(0,4,4)$, $(4,4,4)$, $(7,0,0)$,
$(0,7,0)$, $(0,0,7)$, $(7,7,0)$, $(7,0,7)$, $(0,7,7)$, $(7,7,7)$ (medium-size circles) has three of
them in common; and

\item   each of the remaining 42 triples (small circles) lies on a unique submodule.

\end{itemize}

 It is obvious from Fig.~\ref{fig1} that the seven triples of the
set listed in the second item above can be viewed as the points of the smallest projective
plane, the Fano plane. As these triples are seen to form specif\/ic three-member subsets
each of which def\/ines a unique
aggregate of submodules of cardinality three, and there are just
seven such aggregates (distinguished in Fig.~\ref{fig1} by dif\/ferent colors),
these aggregates of submodules can be regarded as the lines of
this plane. This is a~truly remarkable property because the
entries of the seven triples come from $J$, which is a~ring of
order two {\it without} unity, i.e., the ring not isomorphic to
the Galois f\/ield of two elements.
Turning now to the second maximal ideal ($I_2$), we f\/ind that we
cannot create from it any non-unimodular triple generating free
{\it left} cyclic submodules. If, however, we switch to {\it
right} cyclic submodules instead, we do f\/ind the conf\/iguration
completely identical in its shape to that shown in Fig.~\ref{fig1};
moreover, as $J$, by its very def\/inition, belongs to the both maximal
ideals, the two conf\/igurations share the same Fano plane and, so,
can be regarded as twin (or dual to each other) ``Fano-Snowf\/lakes''.

\begin{figure}[t]
\centerline{\includegraphics[width=5.0truecm,angle=90,clip=]{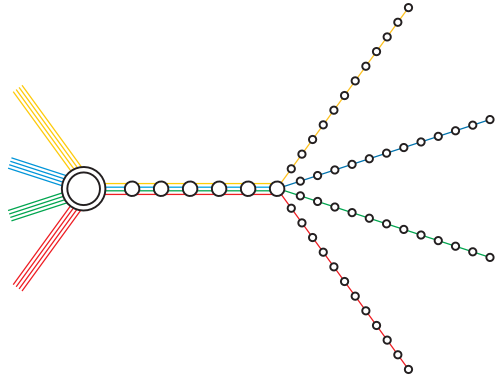}}
\caption{A sketchy illustration -- in the symbols
and notation of Fig.~\ref{fig1} -- of a generic part of the ``Snowf\/lake''
created by free left cyclic submodules generated by non-unimodular
triples of the ring of ternions over $GF(3)$; the double-circle
stands, on the one hand, for {\it two} distinct triples of the
Jacobson radical of the ring and, on the other hand, for  a {\it
single} point of the underlying projective plane of order three.}\label{fig2}
\end{figure}

The Fano plane occurs in algebraic geometry and geometric algebra
in various disguises \cite{pol,br}, and is perhaps most recognized as a
``gadget'' completely describing the algebra structure of the
octonions \cite{baez}. It is, therefore, not surprising that this plane
has also found numerous applications in physics (coding and
information theory, network-switching, etc.). In this context it
is especially worth mentioning recently-discovered relations
between stringy black holes and quantum information theory. We
have in mind intriguing mathematical coincidences between
black-hole solutions in string theory and quantum entanglement in
certain f\/inite-dimensional Hilbert spaces, where some symmetry
structures relevant to string theories are encoded into the
incidence structure of the Fano plane. In particular, dif\/ferent
types of so-called $E_7$-symmetric black-hole solutions can neatly
be classif\/ied in terms of dif\/ferent types of entangled quantum
states attached to the points/lines of the Fano plane and the
black hole entropy formula based on the Fano plane yields an
entanglement measure of seven qubits \cite{l06,l07,df,df2}. It may well happen
that the above-described generalized/extended Fano conf\/igurations,
and a variety of their relatives to be discovered over many
higher-order non-commutative rings, provide a~geo\-metric link
between more involved stringy black holes and more complex forms
of quantum entanglement featured by multipartite quantum systems.
It is already the next case in the hierarchy, viz. the ring of
ternions over the f\/ield of three elements, which deserves a
detailed inspection. Here, in analogy with the Fano case, we f\/ind
a ``snowf\/lake" centered around the projective plane of order three
(see Fig.~\ref{fig2})\footnote{Based on these two cases, we surmise that
the corresponding conf\/iguration always features $PG(2, q)$ as the
core geometry, irrespectively of the order $q$ of the base f\/ield
of the ring of ternions.}.  This projective plane has a noteworthy
link to a few sporadic simple groups \cite{cem}, including the largest
one, the Monster group \cite{cs}. And the Monster is intimately
connected with another wide class of 3D black holes, so-called BTZ
black holes~\cite{btz}, as the logarithm of the dimension of one of its
representations yields the entropy of such a black hole \cite{wit}. It
is, therefore, our hope that this paper will stir the equal
interest of both mathematicians and theoretical physicists: the
former into a systematic treatment of free cyclic submodules
generated by non-unimodular vectors, the latter into a serious
search for possible applications of the associated geometrical
conf\/igurations.

\subsection*{Acknowledgements}

The work was supported by the VEGA grant agency projects
Nos. 6070 and 7012, the CNRS-SAV Project No. 20246
``Projective and Related Geometries for Quantum Information'' and
by the Action Austria-Slovakia project No. 58s2 ``Finite
Geometries Behind Hilbert Spaces''.

\pdfbookmark[1]{References}{ref}
\LastPageEnding


\begin{thebibliography}{99}

\footnotesize\itemsep=0pt

\bibitem{tv}
T\"orner G., Veldkamp F.D., Literature on geometry over rings, {\it J. Geom.} {\bf 42} (1991), 180--200.

\bibitem{psk}
Planat M., Saniga M., Kibler M.R., Quantum entanglement and projective  ring
     geometry, {\it SIGMA}  {\bf 2} (2006), 066, 14~pages, \href{http://arxiv.org/abs/quant-ph/0605239}{quant-ph/0605239}.


\bibitem{spph}
Saniga M., Planat M., Pracna P., Havlicek H., The Veldkamp space of two-qubits,
     {\it SIGMA} {\bf 3} (2007), 075, 7~pages, \href{http://arxiv.org/abs/0704.0495}{arXiv:0704.0495}.

\bibitem{hs07}
Havlicek H., Saniga M., Projective ring line of a specif\/ic qudit, {\it J. Phys.~A:
     Math. Theor.} {\bf 40} (2007), F943--F952, \href{http://arxiv.org/abs/0708.4333}{arXiv:0708.4333}.

\bibitem{pb}
Planat M., Baboin A.-C., Qudits of composite dimension, mutually unbiased bases
     and projective ring geometry, {\it J. Phys. A: Math. Theor.} {\bf 40} (2007), F1005--F1012, \href{http://arxiv.org/abs/0709.2623}{arXiv:0709.2623}.

\bibitem{hs08}
Havlicek H., Saniga M., Projective ring line of an arbitrary single qudit, {\it J.
     Phys. A: Math. Theor.} {\bf 41} (2008), 015302, 12~pages, \href{http://arxiv.org/abs/0710.0941}{arXiv:0710.0941}.

\bibitem{spp}
Saniga M., Planat M., Pracna P., Projective ring line encompassing two-qubits,
     {\it Theor. and Math. Phys.} {\bf  155} (2008), 463--473, \href{http://arxiv.org/abs/quant-ph/0611063}{quant-ph/0611063}.

\bibitem{san}
Saniga M., A f\/ine structure of f\/inite projective ring lines, an invited talk given at the
     workshop on Prolegomena for Quantum Computing (November
     21--22, 2007, Besan\c{c}on, France),  slides of the talk are available at \url{http://hal.archives-ouvertes.fr/hal-00199008}.

\bibitem{v81}
Veldkamp F.D., Projective planes over rings of stable rang 2, {\it Geom. Dedicata} {\bf 11} (1981), 285--308.

\bibitem{v95}
Veldkamp F.D., Geometry over rings, in Handbook of Incidence Geometry, Editor F. Buekenhout, Elsevier, Amsterdam, 1995, 1033--1084.

\bibitem{her}
Herzer A., Chain geometries, in Handbook of Incidence Geometry, Editor F. Buekenhout, Elsevier, Amsterdam, 1995, 781--842.

\bibitem{benz}
Benz W., Zur Umkehrung von Matrizen im Bereich der Ternionen,
{\it Mitt. Math. Ges. Hamburg} {\bf 10} (1979), 509--512.

\bibitem{lpw}
Lex W., Poneleit V., Weinert H.J.,
\" Uber die Einzigkeit der Ternionenalgebra und linksalternative Algebren kleinen Ranges,
{\it Acta Math. Acad. Sci. Hungar.} {\bf 35} (1980), 129--138.

\bibitem{noe}
N\" obauer C., The book of the rings -- part I (2000), pages 65 and 76, available online from
      \mbox{\url{http://www.algebra.uni-linz.ac.at/~noebsi/pub/rings.ps}}.

\bibitem{pol}
Polster B., A geometrical picture book, Springer, New York, 1998, Chapter~5.

\bibitem{br}
Brown E., The many names of (7,3,1), {\it Math. Mag.} {\bf 75} (2002), 83--94.

\bibitem{baez}
Baez J., The octonions, {\it Bull. Amer. Math. Soc. (N.S.)} {\bf
      39} (2002), 145--205, \href{http://arxiv.org/abs/math.RA/0105155}{math.RA/0105155}.

\bibitem{l06}
L\' evay P., Stringy black holes and the geometry of entanglement, {\it Phys. Rev.~D}
      {\bf 74} (2006), 024030, 16~pages, \href{http://arxiv.org/abs/hep-th/0603136}{hep-th/0603136}.

\bibitem{l07}
L\' evay P., Strings, black holes, the tripartite entanglement of seven qubits and the Fano
      plane, {\it Phys. Rev.~D} {\bf 75} (2007), 024024, 19~pages, \href{http://arxiv.org/abs/hep-th/0610314}{hep-th/0610314}.

\bibitem{df}
Duf\/f M.J., Ferrara S., $E_6$ and the bipartite entanglement of three qutrits, {\it Phys.
      Rev. D} {\bf 76} (2007), 124023, 7~pages, \href{http://arxiv.org/abs/0704.0507}{arXiv:0704.0507}.

\bibitem{df2}
Duf\/f M.J., Ferrara S., $E_7$ and the tripartite entanglement of seven qubits,
{\it Phys. Rev. D} {\bf 76} (2007), 025018, 7~pages, \href{http://arxiv.org/abs/quant-ph/0609227}{quant-ph/0609227}.

\bibitem{cem}
Conway J.H., Elkies N.D., Martin J.L., The Mathieu group $M_{12}$ and its pseudogroup extension $M_{13}$,
{\it Expe\-riment. Math.} {\bf 15} (2006), 223--236,
\href{http://arxiv.org/abs/math.GR/0508630}{math.GR/0508630}.

\bibitem{cs}
Conway J.H., Simons C.S., 26 implies the bimonster, {\it J. Algebra}
{\bf 235} (2001), 805--814.

\bibitem{btz}
Ba\~{n}ados M., Teitelboim C., Zanelli J., The black hole in three dimensional space time, {\it Phys. Rev. Lett.}
{\bf 69} (1992), 1849--1851, \href{http://arxiv.org/abs/hep-th/9204099}{hep-th/9204099}.

\bibitem{wit}
Witten E., Three-dimensional gravity revisited,  \href{http://arxiv.org/abs/0706.3359}{arXiv:0706.3359}.

\end{thebibliography}
\end{document}